# Об эффекте кристаллической структуры в распылении двухкомпонентных монокристаллических мишеней

**© 2018 г.   В. Н. Самойлов\*, Н. Г. Ананьева**

*Московский государственный университет имени М.В. Ломоносова,*

*119991, Москва, Россия*

*\*E-mail: samoilov@polly.phys.msu.ru*

С помощью компьютерного моделирования методом молекулярной динамики исследовано распыление на прострел атомов компонент при ионной бомбардировке грани (0001) двухкомпонентного монокристалла $VSi_2$. Рассчитано распыление атомов компонент при распылении виртуальных монокристаллов $VV'_2$ и $Si'Si_2$, состоявших из атомов одной массы, и исследован новый эффект в селективном распылении: при равенстве масс и энергий связи атомов компонент на прострел преимущественно распылялись атомы из ванадиевых узлов. Этот эффект можно назвать эффектом неидентичности узлов компонент в сложной решетке $VSi_2$ (типа C40) к распространению импульса в каскадах столкновений, то есть эффектом структуры в селективном распылении двухкомпонентных монокристаллических мишеней.

**Ключевые слова:** распыление, прохождение ионов, селективное распыление, распыление на прострел, двухкомпонентные мишени, $VSi_2$, тонкие монокристаллические пленки.

## ВВЕДЕНИЕ

При распылении многокомпонентных материалов ионами при малых дозах облучения наблюдается селективное распыление атомов компонент. Коэффициенты распыления атомов компонент оказываются не пропорциональны их концентрации на поверхности, и атомы одной из компонент распыляются преимущественно. Для анализа селективного распыления двухкомпонентных мишеней часто используется формула Андерсена–Зигмунда [1]:

$$\frac{Y_1}{Y_2} = \frac{C_1}{C_2}\left(\frac{M_2}{M_1}\right)^{2m}\left(\frac{U_2}{U_1}\right)^{1-2m}, \tag{1}$$

где $Y_i$ – коэффициент распыления атомов $i$-ой компоненты, $U_i$ – энергия связи атомов $i$-ой компоненты, $C_i$ – концентрация атомов $i$-ой компоненты на поверхности, $M_i$ – масса атомов $i$-ой компоненты. Здесь $m$ – параметр крутизны потенциала взаимодействия, $m = 0.165$ соответствует "универсальному" потенциалу, успешно применявшемуся во многих исследованиях.

Эта аналитическая формула применима только для аморфных твердых тел. В формулу входит неоднозначно определенный подгоночный параметр $m$. Кроме того, выявлены зависимости преимущественного распыления атомов одной из компонент от других параметров, например, от энергии, массы и угла падения бомбардирующих ионов, кристаллической структуры мишени (например, [2–5]).

Обзор экспериментов дает удовлетворительное согласие с формулой Андерсена–Зигмунда в отношении энергий связи атомов компонент и неоднозначные результаты в отношении их масс [6–8].

Для разделения влияния масс и энергий связи атомов компонент были проведены исследования распыления мишеней, состоящих из смеси изотопов (энергии связи для них считаются одинаковыми). Экспериментальные исследования селективного распыления изотопов при малых дозах облучения показали преимущественность эмиссии атомов с меньшей массой [9, 10]. В работе [11] на основе разработанной аналитической модели был исследован поверхностный механизм преимущественного распыления атомов более легкого изотопа, принимая во внимание то, что атомы легкого изотопа более сильно отклоняются в направлении нормали к поверхности при их рассеянии на окружающих атомах более тяжелого изотопа при эмиссии с поверхности.

Однако, как правило, массы изотопов меняются незначительно. Рядом исследователей были созданы модели виртуальных веществ и сплавов (например, [2, 3] и [12, 13]). В работе [12] было исследовано распыление монокристалла меди (изменяя потенциал взаимодействия атом–атом) и распыление систем с различными массами (использовались кислород, медь, золото и их "сплавы" в верхнем атомном слое монокристалла) при постоянной энергии связи. Аналогичные компьютерные эксперименты проводились в работе [13]: брались атомы меди и атомы меди с двойной массой. Энергия связи изменялась от 50 до 200%. В обеих работах подчеркивалась сложная зависимость селективного распыления от масс атомов мишени, ее отличие от теоретической формулы [1]: от массы компонент мишени не обнаружено однозначной связи. Может наблюдаться преимущественное распыление атомов как легкой, так и тяжелой компоненты. От энергии связи наблюдалась приблизительно обратно пропор-циональная зависимость. В работах [2, 3] моделировалось распыление виртуальной аморфной мишени, состоявшей из двух изотопов молибдена с массами 50 и 100 а.е.м.

В работе [4] по модели молекулярной динамики было рассчитано распыление назад и на прострел атомов компонент при ионной бомбардировке грани (0001) монокристалла $VSi_2$. В этой работе было рассчитано распыление атомов компонент при распылении виртуальных монокристаллов $VV'_2$ и $Si'Si_2$, состоявших из атомов одной массы, и был обнаружен новый эффект в селективном распылении: при равенстве масс и энергий связи атомов компонент как назад, так и на прострел преимущественно распылялись атомы из ванадиевых узлов. Этот эффект можно назвать эффектом неидентичности узлов компонент в сложной решетке $VSi_2$ (типа С40) к распространению импульса в каскадах столкновений, то есть эффектом структуры в селективном распылении двухкомпонентных монокристаллических мишеней. Обнаруженный эффект исследовался и в работе [14].

В работах [5, 15] по модели молекулярной динамики была исследована роль поверхностного механизма (то есть рассеяния распыленных атомов компонент на окружающих атомах поверхности при их эмиссии с поверхности) в селективном распылении монокристаллов. Было показано, что в основе поверхностного механизма преимущественного распыления атомов легкой компоненты лежит более сильное отклонение атомов легкой компоненты в направлении нормали к поверхности при их рассеянии на окружающих атомах тяжелой компоненты при эмиссии с поверхности.

В работах [16–18] была исследована разница в энергоспектрах атомов компонент, распыленных с поверхности двухкомпонентных кристаллов $WSi$, $MoSi$ и $VSi$, возникающая вследствие рассеяния распыленных атомов компонент на окружающих атомах поверхности при их эмиссии с поверхности. Было показано, что на стадии эмиссии происходит сильное перераспределение атомов компонент по энергии, которое приводит к появлению особенностей энергоспектров атомов компонент, наблюдаемых экспериментально, в частности, максимума энергоспектра распыленных атомов легкой компоненты, смещенного в сторону меньших энергий по сравнению с положением максимума энергоспектра распыленных атомов тяжелой компоненты [19].

Был выявлен большой вклад поверхностного механизма, то есть столкновений распыленных атомов с окружающими атомами поверхности в процессе эмиссии, в формирование низкоэнергетической части энергоспектров распыленных атомов компонент и обнаружена "двухконусная" структура вылета распыленных атомов низких энергий [16–18].

В последние годы интерес к эффектам, протекающим при ионной бомбардировке двухкомпонентных мишеней, значительно возрос. При этом роль моделирования в выяснении механизмов, определяющих особенности, наблюдаемые в экспериментах по ионной бомбардировке двухкомпонентных мишеней, оказывается весьма значительной. Можно отметить недавние успешные исследования сегрегации [20–24], попытку снижения преимущественной эмиссии атомов кислорода из двухкомпонентной мишени [25], компьютерные исследования методом молекулярной динамики разупорядочивания в многокомпонентных сплавах под действием ионной бомбардировки [26, 27].

В цели настоящей работы входило исследование зависимости селективного распыления при ионной бомбардировке двухкомпонентной монокристаллической мишени от массы атомов мишени по модели, исключавшей влияние разницы энергий связи на коэффициент распыления атомов компонент, а также продолжение исследования эффекта структуры [4, 14] в селективном распылении атомов компонент.

## МОДЕЛЬ РАСЧЕТА

Модель расчета была описана в работах [4, 5, 28]. Здесь мы лишь кратко приводим ее особенности.

Рассчитаны характеристики прохождения ионов и распыления на прострел монокристаллической пленки дисилицида ванадия $VSi_2$, проведено их сравнение с аналогичными характеристиками прохождения ионов и распыления на прострел при ионной бомбардировке виртуальных монокристаллических пленок $VV'_2$ и $Si'Si_2$. Причем кристаллическая решетка для чистых $VV'_2$ и $Si'Si_2$ была взята та же, что и у $VSi_2$, аналогично [4]. Это дало возможность исследовать зависимость распыления не только от массы атомов, но и выявить роль кристаллической структуры в селективном распылении.

Распыление твердых тел под действием ионной бомбардировки можно рассматривать как классическую задачу взаимодействия многих тел. Записав уравнения движения $N$ точечных частиц с массами $m_i$ из второго закона Ньютона, получим систему $6N$ дифференциальных уравнений первого порядка с начальными условиями [29]:

$$m\mathrm{d}v_{ki}/\mathrm{d}t = F_{ki},$$
$$\mathrm{d}x_{ki}/\mathrm{d}t = v_{ki},$$

(2)

где $F_i$ – равнодействующая всех сил, действующих на частицу с номером $i$ со стороны всех остальных частиц (атомов и иона), $i = 1, \ldots, N$; $k = 1, 2, 3$. Определение этих сил – самостоятельная сложная задача. Для упрощения задачи воспользуемся приближениями: силы парные, центральные, потенциальные. В ряде задач физики часто применяют потенциалы межатомного взаимодействия, параметры которых определяются подгонкой под какие-либо экспериментальные данные (используются постоянная решетки, коэффициент теплового расширения, энергия сублимации и т.д.): потенциалы Морзе, Леннарда–Джонса, Борна–Майера и другие.

Подобные системы уравнений решаются численно с использованием разностных схем. Выбор разностной схемы зависит (в том числе) и от приближений сил взаимодействия. Для многочастичного динамического взаимодействия часто используют метод Рунге–Кутта второго порядка [30].

Расчет проводился методом молекулярной динамики. Ионы криптона $Kr^+$ бомбардировали три ультратонкие монокристаллические пленки: $VSi_2$ и виртуальные $VV'_2$ и $Si'Si_2$ с такой же кристаллической структурой, что и $VSi_2$. Более подробно опишем виртуальные кристаллы и поясним обозначения $VV'_2$ и $Si'Si_2$. Виртуальный кристалл $VV'_2$ состоял из атомов V, расположенных в своих узлах решетки, и атомов V' в кремниевых узлах решетки. При этом атомы V' не отличались от атомов V ни по массе, ни по значению энергии связи на поверхности. Таким образом, обозначение V' обозначало те же атомы V, но расположенные до начала облучения ионами в кремниевых узлах решетки. При этом в расчетах мы различали коэффициенты распыления на прострел атомов V и V'.

Аналогично, виртуальный кристалл $Si'Si_2$ состоял из атомов Si, расположенных в своих узлах решетки, и атомов Si' в ванадиевых узлах решетки. При этом атомы Si' не отличались от атомов Si ни по массе, ни по значению энергии связи на поверхности. Таким образом, обозначение Si' обозначало те же атомы Si, но расположенные до начала облучения ионами в ванадиевых узлах решетки. В расчетах мы различали коэффициенты распыления на прострел атомов Si и Si'.

Каждый из кристаллитов состоял из 47 атомов, расположенных в 3-х атомных слоях, параллельных поверхности (рис. 1). Сравнение результатов расчетов для мишеней из 397-ми и 47-ми атомов было проведено в работе [31]. Ионы $Kr^+$ падали перпендикулярно поверхности грани (0001), их энергия $E_0$ изменялась от 50 эВ до 100 кэВ. На каждый кристаллит при каждой энергии $E_0$ падало по 1051-му иону.

Реальные силы взаимодействия атомов (отталкивания–притяжения) заменялись силами отталкивания, а притяжение к кристаллу моделировалось сферическим потенциальным барьером при эмиссии атома с поверхности. В качестве потенциала взаимодействия атом–атом и ион–атом использовался потенциал Борна–Майера, гладко сопряженный с обратноквадратичным потенциалом, аналогичный составному потенциалу [32]. Использование потенциала отталкивания вместо потенциала отталкивания–притяжения дает возможность разделить вклады каскадов столкновений и энергии связи в характеристики распыления. Для интегрирования уравнений движения использовался метод "средней силы" (метод Рунге–Кутта второго порядка). Используя одинаковый потенциальный барьер для ванадия и кремния высотой 4.64 эВ, равной энергии сублимации для кремния [33], мы исключаем влияние разницы энергий связи атомов компонент на селективное распыление.

Рассматривалось прохождение ионов криптона и распыление на прострел атомов компонент, причем отдельно из ванадиевых и кремниевых узлов решетки.

## РЕЗУЛЬТАТЫ И ИХ ОБСУЖДЕНИЕ

**Прохождение ионов через тонкие монокристаллические пленки.** На рис. 2 представлены зависимости коэффициентов прохождения ионов криптона $Kr^+$ на прострел от энергии $E_0$. Для всех мишеней наблюдается монотонный рост коэффициента прохождения ионов с ростом энергии $E_0$. При фиксированной энергии $E_0$ коэффициент прохождения ионов через кристалл Si'Si$_2$ больше, чем через другие образцы, через кристалл VV'$_2$ – меньше, чем через остальные образцы. Это объясняется различием сечений взаимодействия $Kr^+$–Si и $Kr^+$–V. Кроме того, передача энергии от $Kr^+$ атомам Si меньше, чем атомам V вследствие большей разницы масс. Поэтому при взаимодействии с атомами V ионы $Kr^+$ теряют большую долю своей энергии, чем при взаимодействии с атомами Si. В связи с этим, чем больше в мишени ванадия, тем больше вероятность для иона криптона потерять энергию.

Очевидно, вероятность прохождения ионов должна зависеть от концентрации компонент в мишени, что мы и наблюдаем: число прошедших ионов, нормированное на число бомбардирующих ионов, то есть коэффициент прохождения ионов на прострел $R_\downarrow$ в области энергий $E_0$ 100–300 эВ зависит от концентрации компонент в мишени приблизительно как

$$3R_{\downarrow VSi_2} = 2R_{\downarrow Si'Si_2} + R_{\downarrow VV'_2} \qquad (3)$$

(табл.). При увеличении начальной энергии ионов наблюдается отклонение от линейности коэффициента прохождения ионов. При больших энергиях сечение взаимодействия стремится к нулю, а коэффициент прохождения ионов, соответственно, к 1.0. В области малых энер-гий $E_0$ наблюдается особенность, заключающаяся в том, что коэффициент прохождения ионов через кристалл $VSi_2$ оказывается больше, чем через кристалл $Si'Si_2$.

**Распыление на прострел атомов компонент.** На рис. 3 представлены коэффициенты распыления на прострел атомов из разных узлов, нормированные на стехиометрическое отношение (отношение концентраций компонент) для необлученного образца $VSi_2$ (1:2). Из образца виртуального кремния $Si'Si_2$ при всех энергиях ионов $Kr^+$ суммарно распыляется на прострел атомов меньше, чем из других образцов. Это можно объяснить различием сечений взаимодействия и коэффициентов передачи энергии $Kr^+$–$Si$ и $Kr^+$–$V$. Из образца виртуального ванадия $VV'_2$ при энергиях бомбардирующих ионов $E_0 > 100$ эВ суммарно распыляется на прострел наибольшее количество атомов, как и следовало ожидать. При малых энергиях бомбардирующих ионов из $VV'_2$ распыляется на прострел меньше атомов, чем из дисилицида ванадия $VSi_2$. Однако больше атомов распыляется назад, и, в целом, из виртуального ванадия $VV'_2$ при всех энергиях $E_0$ эмитируется больше атомов, чем из других образцов. То, что из $VV'_2$ суммарно распыляется на прострел меньше атомов при низких энергиях бомбардирующих ионов, говорит о большей "плотности" образца: частицы чаще сталкиваются и успевают развернуться и вылететь назад. По этой же причине максимум энергоспектра атомов, распыленных на прострел из образца $VV'_2$, смещен в сторону больших энергий.

Прозрачность кристаллита (при высоких энергиях $E_0$) возрастает с переходом от $VV'_2$ к $VSi_2$ и $Si'Si_2$. С переходом от $Si'Si_2$ к $VSi_2$ среднее сечение взаимодействия ионов $Kr^+$ с атомами мишени возрастает, значительное количество ионов отдает практически всю энергию атомам кристалла. Это, естественно, приводит к возрастанию суммарного коэффициента распыления на прострел. Хотя число настоящих атомов кремния $Si$ (расположенных в своих узлах решетки) в кристалле не изменилось, распыление на прострел таких атомов кремния возросло. Очевидно, это связано с большим сечением взаимодействия $Kr^+$–$V$ (по сравнению с сечением взаимодействия $Kr^+$–$Si'$), возникновением большего числа смещенных атомов ванадия $V$ (по сравнению с числом смещенных атомов $Si'$) и последующим перераспределением энергии в каскаде столкновений.

Из-за ограниченности облучаемого образца концентрация ванадия и кремния по слоям атомов была различна. В верхнем слое атомов ванадия несколько больше, чем в отношении 1:2 (0.58 вместо 0.5). Второй и третий слои обеднены ванадием (по 0.4 вместо 0.5). В целом, в образце $VSi_2$ атомов ванадия несколько меньше нормы (0.47 вместо 0.5). Тем не менее, наблюдалось преимущественное распыление на прострел атомов из ванадиевых узлов для всех трех образцов, даже когда во всех узлах располагались атомы с одинаковой массой и энергией связи.

На рис. 4 представлены отношения коэффициентов распыления атомов на прострел из ванадиевых и кремниевых узлов в зависимости от начальной энергии $E_0$ при бомбардировке кристаллов $VSi_2$, $Si'Si_2$ и $VV'_2$ ионами $Kr^+$. При низких энергиях большую роль играет разница сечений взаимодействия атомов. Если рассматривать образец виртуального ванадия, то в кремниевом узле атом $V'$ просто не должен помещаться. Однако, за время развития каскадов столкновений и распыления сам он (без дополнительного ускорения) не успевает вылететь из блока атомов (области взаимодействия). Естественно ожидать, что из кремниевых узлов ванадий должен распыляться на прострел преимущественно. Это и наблюдается при низких энергиях $E_0$ бомбардирующих ионов. Однако при энергиях $E_0 \geq 200$ эВ наблюдается преимущественное распыление на прострел атомов ванадия из собственных узлов решетки. Этот результат согласуется с выводами работы [4]. Причем надо отметить, что мы имеем качественное и в значительной степени и количественное соответствие отношения атомов, распыленных из ванадиевых и кремниевых узлов, для всех трех образцов $VV'_2$, $VSi_2$ и $Si'Si_2$, особенно выраженное при энергиях $E_0$, равных 1–10 кэВ.

Связано это, по-видимому, с особенностями распространения каскадов столкновений в монокристаллах сложной структуры (в нашем случае с решеткой типа C40) и разницей передачи импульса по ванадиевой и кремниевой подрешеткам монокристалла. Из этого можно сделать вывод, что пространственная структура мишени играет решающую роль в селективном распылении монокристаллов с решеткой C40 при энергии $E_0 = 1$–10 кэВ, по крайней мере, по сравнению с массой атомов мишени. Тип кристаллической решетки зависит от взаимодействия атомов в кристалле. Поэтому целесообразно искать зависимость селективного распыления от потенциала взаимодействия атомов между собой (энергия связи атомов, которая входит в теоретические формулы оценки селективного распыления, тоже зависит от потенциала взаимодействия атомов).

## ЗАКЛЮЧЕНИЕ

В работе исследованы коэффициенты прохождения ионов и коэффициенты распыления на прострел атомов компонент при ионной бомбардировке ионами $Kr^+$ кристаллитов $VSi_2$, $VV'_2$ и $Si'Si_2$.

Обнаружены и исследованы особенности коэффициента прохождения ионов $Kr^+$ на прострел в зависимости от начальной энергии ионов и "плотности" мишени. При переходе к более "плотной" мишени VV'$_2$, как правило, наблюдается уменьшение коэффициента прохождения ионов.

Расчеты показали, что при не слишком малых энергиях бомбардирующих ионов наблюдается преимущественное распыление на прострел атомов из ванадиевых узлов. Это может быть связано с неидентичностью ванадиевых узлов по отношению к распылению (на прострел). Механизм распыления атомов из ванадиевых узлов может быть отличен от механизма распыления атомов из кремниевых узлов в силу сложной структуры монокристалла VSi$_2$ (типа C40) и разных механизмов распространения импульса по ванадиевой и кремниевой подрешеткам в VSi$_2$.

Очевидно, что этот эффект неидентичности узлов VSi$_2$ по отношению к распылению нужно учитывать при анализе элементного состава многокомпонентных монокристаллов методами ВИМС и МСВИ.

# СПИСОК ЛИТЕРАТУРЫ

# The Effect of Crystal Structure in Sputtering
# of Two-component Single Crystal Targets

## V. N. Samoilov,  N. G. Ananieva

In the present paper the forward sputtering of atoms of the components under ion bombardment of the single crystal $VSi_2$ (0001) face is studied with the use of molecular dynamics computer simulations. Sputtering of atoms of the components under ion bombardment of the virtual single crystals $VV'_2$ and $Si'Si_2$ consisting of atoms of the same mass but located in the lattice sites of V and Si sublattices is calculated. The new effect in the selective sputtering of two-component targets is studied: atoms from the vanadium lattice sites are preferentially forward sputtered for equal masses and equal surface binding energies of atoms of the components. This effect one could name the effect of nonidentity of lattice sites of the components in the complex lattice of $VSi_2$ (of the C40 type) to the propagation of momentum in collision cascades and thus the effect of structure in the selective sputtering of two-component single crystal targets.

**Key words:** sputtering, transmission of ions, selective sputtering, transmission sputtering, two-component targets, $VSi_2$, thin single-crystal films.

## ПОДПИСИ К РИСУНКАМ

Рис. 1. Расположение атомов монокристалла $VSi_2$ (решетка С40) в проекции на поверхность грани (0001). Показан кристаллит из 47-ми атомов, расположенных в трех слоях, параллельных поверхности. Формула чередования слоев АВС, где А – атомный слой на поверхности, атомы показаны черным цветом, второй атомный слой В – серым, третий атомный слой С – белым.

Рис. 2. Зависимости коэффициентов прохождения ионов $Kr^+$ на прострел от начальной энергии $E_0$ при бомбардировке грани (0001) кристаллов $VSi_2$, $Si'Si_2$ и $VV'_2$.

Рис. 3. Зависимости коэффициентов распыления на прострел атомов из ванадиевых (сплошные линии) и кремниевых (пунктирные линии) узлов от начальной энергии ионов $E_0$ при бомбардировке грани (0001) кристаллов $VSi_2$, $VV'_2$ и $Si'Si_2$ ионами $Kr^+$. Коэффициенты распыления нормированы на стехиометрическое отношение для необлученного образца (1:2).

Рис. 4. Зависимости отношения коэффициентов распыления на прострел атомов из ванадиевых и кремниевых узлов от начальной энергии ионов $E_0$ при бомбардировке грани (0001) кристаллов $VSi_2$, $VV'_2$ и $Si'Si_2$ ионами $Kr^+$. Отношения нормированы на стехиометрическое отношение для необлученного образца (1:2).

**Таблица.** Коэффициент прохождения ионов $Kr^+$ через ультратонкие пленки $VV'_2$, $Si'Si_2$ и $VSi_2$ в зависимости от энергии $E_0$.

| Энергия бомбардирующих ионов $E_0$, эВ | Коэффициент прохождения ионов на прострел $R_\downarrow$ | | |
|---|---|---|---|
| | $VV'_2$ | $Si'Si_2$ | $VSi_2$ |
| 100 | 0.078 | 0.555 | 0.406 |
| 150 | 0.462 | 0.906 | 0.749 |
| 200 | 0.625 | 0.975 | 0.858 |
| 300 | 0.776 | 0.999 | 0.952 |
| 500 | 0.887 | 1.000 | 0.998 |

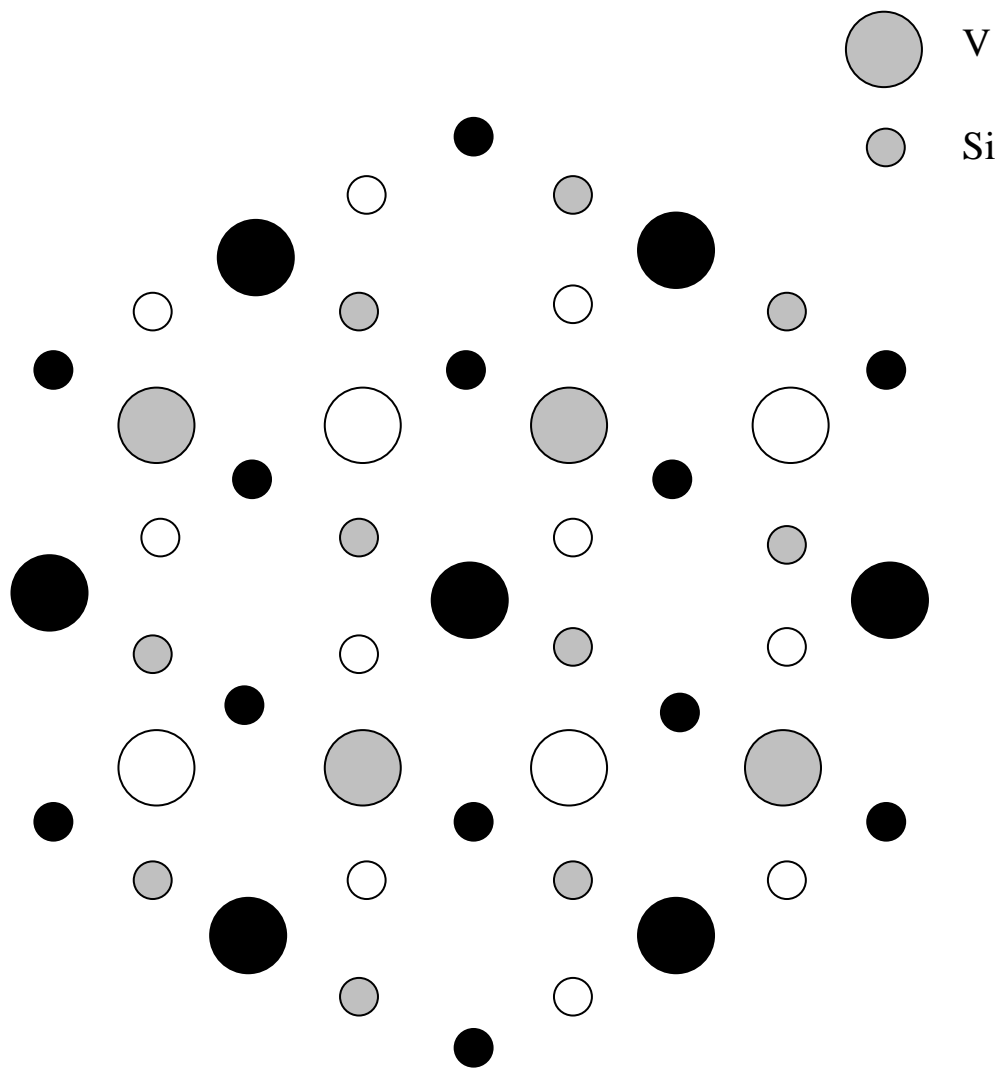

Рис. 1. Расположение атомов монокристалла VSi$_2$ (решетка C40) в проекции на поверхность грани (0001). Показан кристаллит из 47-ми атомов, расположенных в трех слоях, параллельных поверхности. Формула чередования слоев ABC, где A – атомный слой на поверхности, атомы показаны черным цветом, второй атомный слой B – серым, третий атомный слой C – белым.

В. Н. Самойлов, Н. Г. Ананьева

ОБ ЭФФЕКТЕ КРИСТАЛЛИЧЕСКОЙ СТРУКТУРЫ В РАСПЫЛЕНИИ ДВУХКОМПОНЕНТНЫХ МОНОКРИСТАЛЛИЧЕСКИХ МИШЕНЕЙ

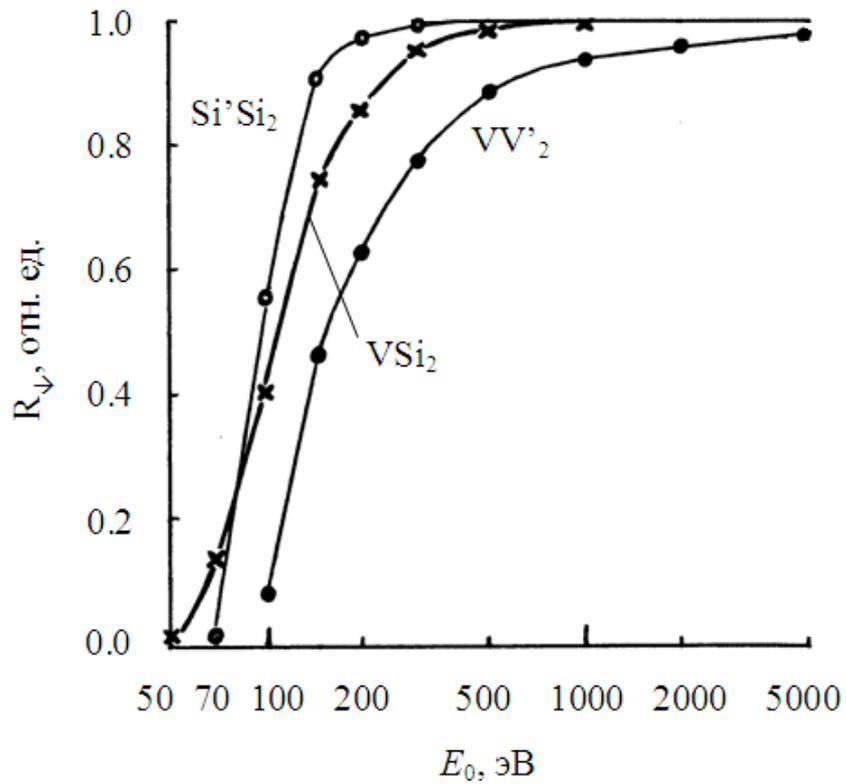

Рис. 2. Зависимости коэффициентов прохождения ионов $Kr^+$ на прострел от начальной энергии $E_0$ при бомбардировке грани (0001) кристаллов $VSi_2$, $Si'Si_2$ и $VV'_2$.

В. Н. Самойлов, Н. Г. Ананьева

ОБ ЭФФЕКТЕ КРИСТАЛЛИЧЕСКОЙ СТРУКТУРЫ В РАСПЫЛЕНИИ
ДВУХКОМПОНЕНТНЫХ МОНОКРИСТАЛЛИЧЕСКИХ МИШЕНЕЙ

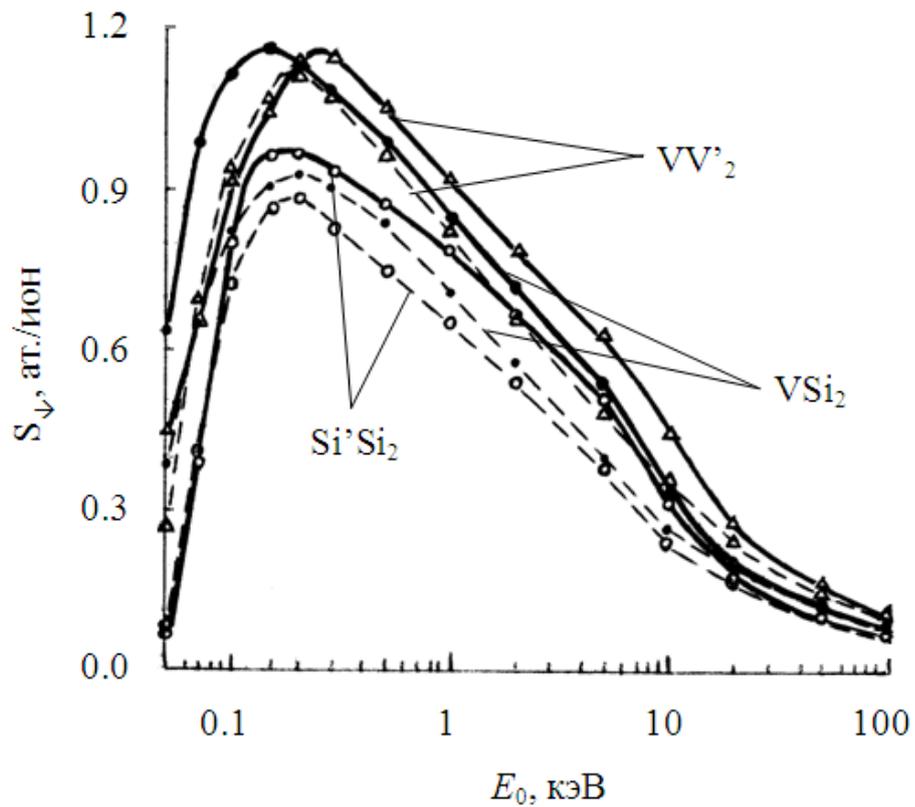

Рис. 3. Зависимости коэффициентов распыления на прострел атомов из ванадиевых (сплошные линии) и кремниевых (пунктирные линии) узлов от начальной энергии ионов $E_0$ при бомбардировке грани (0001) кристаллов $VSi_2$, $VV'_2$ и $Si'Si_2$ ионами $Kr^+$. Коэффициенты распыления нормированы на стехиометрическое отношение для необлученного образца (1:2).

В. Н. Самойлов, Н. Г. Ананьева

ОБ ЭФФЕКТЕ КРИСТАЛЛИЧЕСКОЙ СТРУКТУРЫ В РАСПЫЛЕНИИ
ДВУХКОМПОНЕНТНЫХ МОНОКРИСТАЛЛИЧЕСКИХ МИШЕНЕЙ

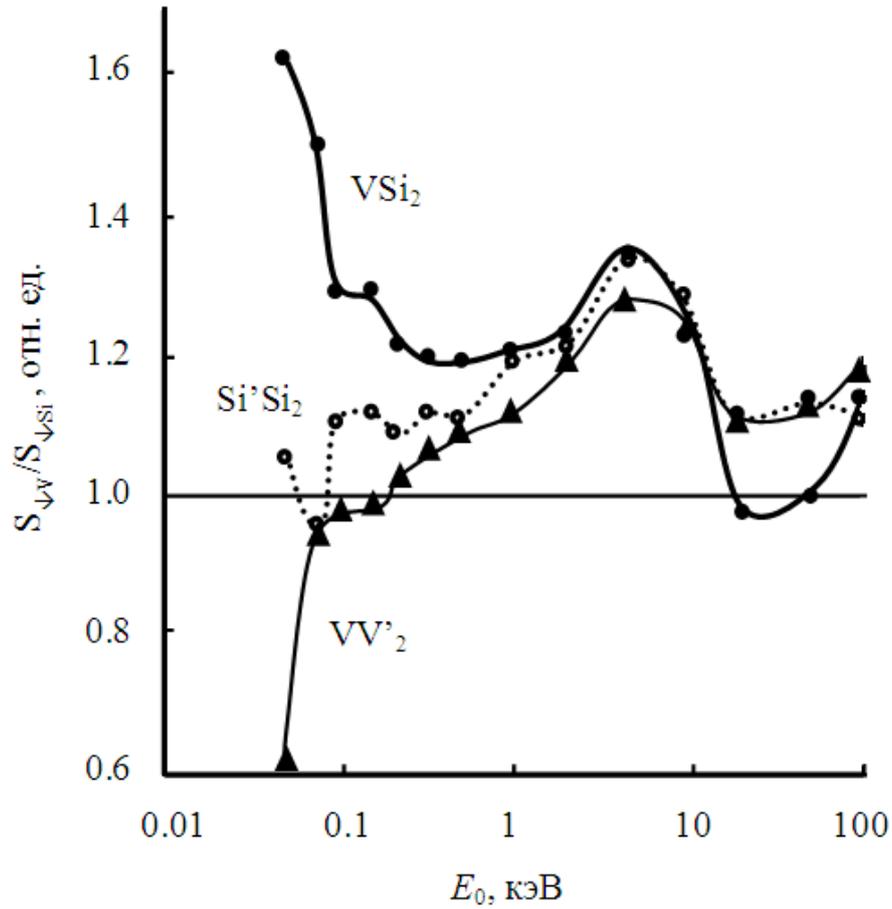

Рис. 4. Зависимости отношения коэффициентов распыления на прострел атомов из ванадиевых и кремниевых узлов от начальной энергии ионов $E_0$ при бомбардировке грани (0001) кристаллов $VSi_2$, $VV'_2$ и $Si'Si_2$ ионами $Kr^+$. Отношения нормированы на стехиометрическое отношение для необлученного образца (1:2).

В. Н. Самойлов, Н. Г. Ананьева

ОБ ЭФФЕКТЕ КРИСТАЛЛИЧЕСКОЙ СТРУКТУРЫ В РАСПЫЛЕНИИ
ДВУХКОМПОНЕНТНЫХ МОНОКРИСТАЛЛИЧЕСКИХ МИШЕНЕЙ